\documentclass[aps,twocolumn,longbibliography] {revtex4-1}
\usepackage{amsmath,graphicx,amsfonts}

\usepackage{times}
\usepackage[colorlinks=true,linkcolor=blue,urlcolor=blue,citecolor=blue, breaklinks=true,hypertexnames=false]{hyperref}

\usepackage{amssymb}
\usepackage{bbm}

\usepackage[utf8]{inputenc}
\DeclareUnicodeCharacter{03BC}{\mbox{$\mu$}}

\DeclareUnicodeCharacter{03B4}{$\delta$}
\DeclareUnicodeCharacter{2009}{-}

\usepackage[english]{babel}
\addto\captionsenglish{}
\addto\captionsenglish{}

\makeatletter
\def\bbl@set@language#1{%
	\edef\languagename{%
		\ifnum\escapechar=\expandafter`\string#1\@empty
		\else\string#1\@empty\fi}%
	\@ifundefined{babel@language@alias@\languagename}{}{%
		\edef\languagename{\@nameuse{babel@language@alias@\languagename}}%
	}%
	\select@language{\languagename}%
	\expandafter\ifx\csname date\languagename\endcsname\relax\else
	\if@filesw
	\protected@write\@auxout{}{\string\select@language{\languagename}}%
	\bbl@for\bbl@tempa\BabelContentsFiles{%
		\addtocontents{\bbl@tempa}{\xstring\select@language{\languagename}}}%
	\bbl@usehooks{write}{}%
	\fi
	\fi}
\newcommand{\DeclareLanguageAlias}[2]{%
	\global\@namedef{babel@language@alias@#1}{#2}%
}
\makeatother

\DeclareLanguageAlias{en}{english}

\makeatletter
\def\@bibdataout@aps{%
	\immediate\write\@bibdataout{%
		@CONTROL{%
			apsrev41Control%
			\longbibliography@sw{%
				,author="08",editor="1",pages="1",title="0",year="1"%
			}{%
				,author="08",editor="1",pages="1",title="",year="1"%
			}%
		}%
	}%
	\if@filesw \immediate \write \@auxout {\string \citation {apsrev41Control}}\fi 
}
\makeatother

\date\today

\def\TITLE{Exact Results for the Moments of the Rapidity Distribution in Galilean-Invariant Integrable Models}

\begin{document}
\title{\TITLE}
\author{Zoran Ristivojevic}
\affiliation{Laboratoire de Physique Th\'{e}orique, Universit\'{e} de Toulouse, CNRS, UPS, 31062 Toulouse, France}

\begin{abstract}
We study a class of Galilean-invariant one-dimensional Bethe ansatz solvable models in the thermodynamic limit. Their rapidity distribution obeys an integral equation with a difference kernel over a finite interval, which does not admit a closed-form solution. We develop a general formalism enabling one to study the moments of the rapidity distribution, showing that they satisfy a difference-differential equation. The derived equation is explicitly analyzed in the case of the Lieb-Liniger model and the moments are analytically calculated. In addition, we obtained the exact information about the ground-state energy at weak repulsion. The obtained results directly enter a number of physically relevant quantities.
\end{abstract}
\maketitle

\textit{Introduction.---} Renewed broad interest in quantum integrable systems, beyond the field of mathematical physics, arises from their experimental realizations with cold gases \cite{cazalilla_one_2011,guan_fermi_2013}. One of the main particular features of integrable systems is the existence of an extensive number of conserved quantities opposite to very few ones in generic systems. They strongly constrain the time evolution of an initial state of the system, globally affecting the dynamics and thermalization \cite{rigol_relaxation_2007,kinoshita_quantum_2006}. A natural key question that emerged was how to construct the generalized thermodynamic ensemble in order to describe the stationary state of the system at late times. It is nowadays widely accepted that the conventional Gibbs ensemble for generic systems is replaced by a more general one involving the conserved quantities (so-called charges) \cite{essler_generalized_2015,ilievski_complete_2015}, which has also been supported experimentally \cite{langen_experimental_2015}.

A Bethe ansatz integrable model is characterized by the exact wave function. The latter is parametrized by the set of rapidities that obey the Bethe equations. In the thermodynamic limit, it is appropriate to consider the rapidity distribution. It has a simple physical meaning in the special case of $\delta$-interacting bosons in one dimension (i.e., the Lieb-Liniger model) at infinite repulsion strength. Then the rapidities coincide with the momenta of a free Fermi gas, having thus a constant density. Decreasing the repulsion strength, the set of rapidities evolves according to the Bethe ansatz equations and the distribution shrinks symmetrically. At weak interaction, the distribution becomes sharply peaked around zero momentum, which marks a tendency of bosons to exhibit a Bose-Einstein condensation. Interestingly, initially conceived theoretically, the rapidity distribution has been directly measured in a recent experiment \cite{wilson_observation_2020}.

The rapidity distribution is a central quantity that determines various physically important quantities in integrable models. The ground-state energy is proportional to the second moment of the rapidity distribution. A number of correlation functions have also been expressed in terms of the second and higher moments. Well-known examples include the short-distance expansion of the one-body density matrix  \cite{olshanii_short-distance_2003,olshanii_connection_2017} as well as the local $2$- and $3$-body correlation functions \cite{gangardt_stability_2003,cheianov_exact_2006}. In fact, the latter should be true in the more general $N$-body case. The exponent of the decay of the one-body density matrix is a function of the value of the rapidity distribution at the edge \cite{haldane_effective_1981}. Remarkably, even the spectrum of elementary excitations can be obtained from the rapidity distribution of the system in the ground state \cite{petkovic_spectrum_2018}, which further emphasizes its importance. Finally, the moments are proportional to the expectation values of the conserved charges.

The rapidity distribution is governed by an integral equation, see Eq.~(\ref{eq:F}) below, with unknown closed-form solution. This complicates the evaluation of the moments, which are typically not known analytically, apart from the ground-state energy in some cases. In this paper we make a significant progress in this direction. We develop the formalism for the analytical evaluation of the moments of the rapidity distribution in one-dimensional Galilean-invariant integrable models. We find an exact differential equation for the moment-generating function, which reduces to a difference-differential equation for the moments. The latter is then analyzed on the example of the Lieb-Liniger model and explicit analytical results are obtained.

\textit{General results.---} We consider an integrable many-body system of nonrelativistic quantum particles with the pairwise interactions that depend on the relative coordinate of particles in the thermodynamic limit. In such Galilean-invariant systems, the density of rapidities (or rapidity distribution) obeys the Lieb integral equation \cite{sutherland}
\begin{align}\label{eq:F}
	\rho(k,Q)+\frac{1}{2\pi}\int_{-Q}^{Q} dq\theta'(k-q)\rho(q,Q)=\frac{1}{2\pi}.
\end{align}
Here $Q$ is the Fermi rapidity, which denotes the highest occupied rapidity in the ground state. In Eq.~(\ref{eq:F}), $\theta'(k)$ denotes the derivative of the two-particle scattering phase shift, which is an even real function. This implies that the density of rapidities is an even positive function, $\rho(k,Q)=\rho(-k,Q)$. Differentiating Eq.~(\ref{eq:F}), after using the partial integration and the parity of $\rho(k,Q)$ and $\theta'(k)$, one obtains that the density of rapidities determined by Eq.~(\ref{eq:F}) also satisfies a partial differential equation \cite{Note1}
\begin{align}\label{eq:PDE}
	\left(\frac{\partial^2}{\partial Q^2}-2\frac{d}{dQ}[\ln \rho(Q,Q)] \frac{\partial}{\partial Q} -\frac{\partial^2}{\partial k^2}\right)\rho(k,Q)=0.
\end{align}
Instead of Eq.~(\ref{eq:F}), in the following considerations we will use Eq.~(\ref{eq:PDE}) as a starting point. 
\footnotetext{See the Supplemental Material for the details.}

In order to study the moments of the rapidity distribution, it is useful to consider an integral
\begin{align}\label{eq:f}
	f_\alpha(Q)=\int_{-Q}^{Q} dk \rho(k,Q) \cosh(\alpha k),
\end{align}
where $\alpha$ is a real parameter. Equation (\ref{eq:f}) can be understood as the moment-generating function, since the moments of $\rho(k,Q)$ can be obtained by differentiating $f_\alpha(Q)$ with respect to $\alpha$ and then taking the limit $\alpha\to 0$. The real usefulness of $f_\alpha(Q)$ arises from the relation 
\begin{align}\label{eq:main}
	\left(\frac{\partial^2}{\partial Q^2}-2\frac{d}{dQ}[\ln \rho(Q,Q)] \frac{\partial}{\partial Q}
	\right) f_\alpha(Q)=\alpha^2 f_\alpha(Q),
\end{align}
which can be shown directly by applying the derivatives to the definition (\ref{eq:f}) after making use of Eq.~(\ref{eq:PDE}). Equation (\ref{eq:main}) is an exact result that is derived under the minimal assumption that the scattering phase shift is a repeatedly differentiable function. It thus applies to all integrable models where the density of rapidities is determined by Eq.~(\ref{eq:F}) with smooth $\theta(k)$. Three well-known examples are the hyperbolic Calogero-Sutherland \cite{sutherland}, the Lieb-Liniger \cite{lieb_exact_1963}, and the Yang-Gaudin models \cite{gaudin_2014}. 

Consider the (dimensionless) moments of the rapidity distribution normalized as
\begin{align}\label{eq:e2l}
	e_{2l}=\frac{1}{n^{2l+1}} \int_{-Q}^{Q} dk k^{2l}\rho(k,Q),
\end{align}
where $l\ge 0$ is an integer and $n$ is the density of particles, defined by
$n=f_0(Q)$. Therefore the lowest moment is $e_0=1$, while the higher ones can be obtained from Eq.~(\ref{eq:f}) since $n^{2l+1}e_{2l}=(\partial^{2l} f_\alpha(Q)/\partial\alpha^{2l})|_{\alpha=0}$. Using the relation  $dn/dQ=4\pi\rho^2(Q,Q)$ \cite{Note2}\footnotetext{Equation (\ref{eq:main}) at $\alpha=0$ leads to $dn/dQ=A\rho^2(Q,Q)$, where the integration constant can be set to $A=4\pi$ using the free Fermi gas case. This is in agreement with another derivation presented in Ref.~\cite{korepin1993book}.}\nocite{korepin1993book} to express the derivative in the left-hand side of Eq.~(\ref{eq:main}) as $16\pi^2\rho^{4}(Q,Q)\partial^2/\partial n^2$,
we obtain
\begin{align}\label{eq:main1}
\frac{\partial^2}{\partial n^2}(n^{2l+1}e_{2l})=\frac{l(2l-1)}{8\pi^2\rho^4(Q,Q)} n^{2l-1} e_{2l-2}.	
\end{align}
Equation (\ref{eq:main1}) is the main result of this paper. It shows a remarkable fact that the moments of the rapidity distribution  (\ref{eq:e2l}) are not independent, but must satisfy a difference-differential equation, which is given by Eq.~(\ref{eq:main1}). In the following we study its consequences in more details.

At $l=0$, Eq.~(\ref{eq:main1}) is trivial, while at $l=1$ it leads to
\begin{align}\label{eq:main2}
	\frac{\partial^2}{\partial n^2}(n^{3}e_{2})=\frac{n}{8\pi^2\rho^4(Q,Q)}.
\end{align}
Equation (\ref{eq:main2}) is equivalent to the thermodynamic expression for the velocity of excitations $v$ that is given by $v^2=(L/mn) (\partial^2 E_0/\partial L^2)$. Here $L$ is the system size, $m$ is the mass of particles, and $E_0$ is the ground-state energy, which is related to the second moment via the relation $E_0=\hbar^2 n^3 L\:\! e_2/2m$. At this point we also need the general relation $mvK=\pi\hbar n$ valid for Galilean invariant models, where $K=4\pi^2\rho^2(Q,Q)$ denotes the Luttinger liquid parameter \cite{sutherland,haldane_effective_1981}. For $l\ge 2$, Eq.~(\ref{eq:main1}) uncovers a new set of relations between the moments, enabling us to use the explicit result for one of them to obtain all the others, which we do next.

\textit{Application to the Lieb-Liniger model.---} Previous results do not rely on any specific form of the interaction, but on minimal requirements on the scattering phase shift. Let us now analyze Eq.~(\ref{eq:main1}) in the case of the Lieb-Liniger model. It describes bosons of the mass $m$ interacting via a contact interaction of the strength $\hbar^2 c/m$, and the phase shift is $\theta(k)=-2\arctan(k/c)$. The dimensionless interaction parameter of the model is $\gamma=c/n$ \cite{lieb_exact_1963}. The moments (\ref{eq:e2l}) are dimensionless functions and can be expressed only in terms of $\gamma$. Equation (\ref{eq:main1}) then becomes
\begin{align}\label{eq:e2l1}
		\frac{d^2}{d\gamma^2} \left(\frac{e_{2l+2}}{\gamma^{2l+2}}\right)= (l+1)(2l+1) \frac{d^2}{d\gamma^2} \left(\frac{e_{2}}{\gamma^{2}}\right) \frac{e_{2l}}{\gamma^{2l}}.
\end{align}
For $l=0$, Eq.~(\ref{eq:e2l1}) becomes an identity, while for $l>0$ it enables us to evaluate $e_{2l+2}$ using the known analytical result for $e_2$. This can be achieved analytically in two regimes.

In the regime of weak interactions, $\gamma\ll 1$, the leading-order solution of Eq.~(\ref{eq:F}) is $\rho(k,Q)=\sqrt{Q^2-k^2}/2\pi c$ \cite{lieb_exact_1963}. This gives the order of magnitude estimate for the leading-order term in Eq.~(\ref{eq:e2l}), $e_{2l}\sim\left({Q}/{n}\right)^{2l+2}/\gamma$. Using $e_0=1$, we find $Q\sim n\sqrt{\gamma}$ and thus $e_{2l}(\gamma)\sim \gamma^l$. Since the subsequent terms in the expansion of $e_2$ are multiplied by  $\sqrt{\gamma}$, we assume
\begin{align}\label{eq:e2lass}
	e_{2l}=\sum_{j=0}^{\infty} a_j^{(2l)}\gamma^{l+j/2},
\end{align}
where the values of the numerical coefficients $a_j^{(2l)}$ for $l>1$ will be calculated using the known values of $a_j^{(2)}$ \cite{marino_exact_2019,ristivojevic_conjectures_2019}. Substitution of the form (\ref{eq:e2lass}) into Eq.~(\ref{eq:e2l1}) yields the connection between the coefficients $a_{k}^{(2l+2)}$ from the left-hand side of Eq.~(\ref{eq:e2l1}) and the ones from the right-hand side,
\begin{align}\label{eq:akoff}
	&\left(2l+2-k\right) \left(2l+4-k\right)	a_{k}^{(2l+2)} \notag\\
	&=(l+1)(2l+1) \sum_{j=0}^{k} (j-2)(j-4) a_{j}^{(2)} a_{k-j}^{(2l)}.
\end{align}
Equation~(\ref{eq:akoff}) is trivial for $l=0$ since $a_{k-j}^{(0)}=\delta_{k,j}$, while for $l>1$ it enables us to evaluate the coefficients in the series (\ref{eq:e2lass}) for $e_{2l}$ using the ones of $e_2$. For a fixed $k$, Eq.~(\ref{eq:akoff}) can be explicitly solved since it is equivalent to a first-order linear difference equation \cite{mickens2015difference}. Rather than doing that, in Table \ref{table1} we give the analytical values for  $a_{k}^{(2l)}$ for $1\le l\le 4$. A motivated reader can easily obtain the coefficients for higher values of $l$.

\begin{table}\label{table1}
	\caption{Values of the coefficients in the series (\ref{eq:e2lass}) evaluated from Eqs.~(\ref{eq:akoff}) using the known values of $a_k^{(2)}$.}
	\begin{ruledtabular}
		\begin{tabular}{l|l|l	|l|l}
			$a_k^{(2l)}$ &$k=0$&$k=1$ &$k=2$&$k=3$\\ \hline
			$l=1$ & $1$ & $-\frac{4}{3\pi}$& $\frac{1}{6}-\frac{1}{\pi^2}$ & $-\frac{1}{2\pi^3}+\frac{3\zeta(3)}{8\pi^3}$\\
			$l=2$ & $2$ &$-\frac{88}{15\pi}$ & $1-\frac{2}{\pi^2}$ &$-\frac{4}{3\pi}+\frac{1}{\pi^3}+\frac{21\zeta(3)}{4\pi^3}$ \\
			$l=3$& $5$ & $-\frac{824}{35\pi}$ & $5+\frac{14}{3\pi^2}$ & $-\frac{44}{3\pi}+\frac{17}{\pi^3}+\frac{165\zeta(3)}{4\pi^3}$ \\
			$l=4$ & $14$ & $-\frac{29168}{315\pi}$& $\frac{70}{3}+\frac{3452}{45\pi^2}$ & $-\frac{1648}{15\pi}+\frac{1438}{15\pi^3}+\frac{525\zeta(3)}{2\pi^3}$  \\
		\end{tabular}
	\end{ruledtabular}
\end{table}

In the regime of strong interactions, $\gamma\gg 1$, the integral in the integral operator of Eq.~(\ref{eq:F}) is subdominant and thus $\rho(k,Q)=1/2\pi$ at the leading order. This gives rise to $e_{2l}\sim 1$. Since the subsequent terms in $\rho(k,Q)$ are by a factor of $1/\gamma$ smaller, the resulting series for its moments should be assumed in the form
\begin{align}\label{eq:e2gass}
	e_{2l}=\sum_{j=0}^{\infty} b_j^{(2l)}\gamma^{-j}.
\end{align}
Substituting Eq.~(\ref{eq:e2gass}) into Eq.~(\ref{eq:e2l1}) we find an equation
\begin{align}\label{eq:diffg}
	b_k^{(2l+2)}={}&\frac{(l+1)(2l+1)}{(2l+2+k)(2l+3+k)}\notag\\
	&\times\sum_{j=0}^{k}(2+j)(3+j)b_j^{(2)} b_{k-j}^{(2l)}
\end{align}
that relates the coefficients of Eq.~(\ref{eq:e2gass}). Equation (\ref{eq:diffg}) is a difference equation that has a similar structure as Eq.~(\ref{eq:akoff}), and thus it can be solved for $l>1$. The first five terms are given by
\begin{subequations}
	\label{eeq1}
	\begin{align}
		b_0^{(2l)}={}&\frac{\pi^{2l}}{2l+1},\quad b_1^{(2l)}=-\frac{4l\;\!\pi^{2l}}{2l+1},\quad
		b_2^{(2l)}=4l\pi^{2l},\\ 
		b_3^{(2l)}={}&-\frac{16l(l+1)\pi^{2l}}{3}\biggl[1-\frac{\pi^2}{(2l+1)(2l+3)} \biggr],\\
		b_4^{(2l)}={}&\frac{8l(l+1)(2l+3)\pi^{2l}}{3}\biggl[1-\frac{4\pi^2}{(2l+1)(2l+3)}\biggr].\!
	\end{align}
\end{subequations}
Here we have used the known values of $b_j^{(2)}$ entering $e_2$ \cite{ristivojevic_excitation_2014}, which can be recovered from Eqs.~(\ref{eeq1}) setting $l=1$. The results given in Table~\ref{table1} substituted in Eq.~(\ref{eq:e2lass}) and the ones of Eq.~(\ref{eeq1}) substituted in Eq.~(\ref{eq:e2gass}) illustrate how the exact relation (\ref{eq:e2l1}) for the Lieb-Liniger model can be used to obtain analytically \textit{all} the moments of the rapidity distribution in both regimes of weak and strong interaction from the knowledge of $e_2$ only. 

\begin{figure}
	\includegraphics[width=0.95\columnwidth]{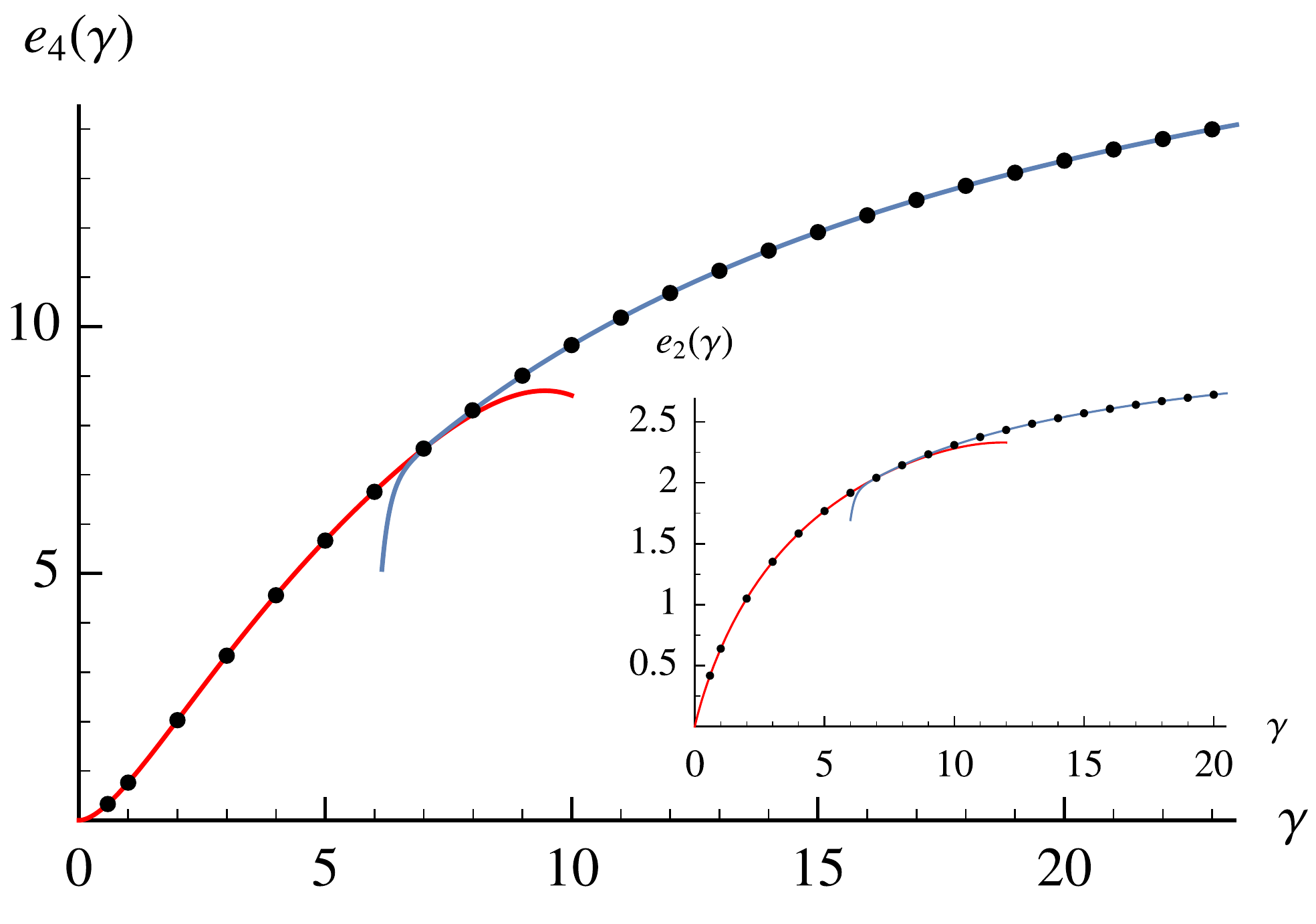}
	\caption{The fourth moment of the rapidity distribution $e_4(\gamma)$ as a function of the interaction strength $\gamma$. The dots represent numerically  exact values; the two curves are obtained from the asymptotic series (\ref{eq:e2lass}) with 11 terms \cite{ristivojevic_method_2022}  and the series (\ref{eq:e2gass}) with 39 terms. The former low-$\gamma$ series agrees well with the exact values for $\gamma\lesssim 7$ (the absolute value of the relative error is $0.01$ at $\gamma=7$, becoming progressively smaller at smaller $\gamma$); the latter high-$\gamma$ series applies for $\gamma\gtrsim 7$ (the absolute value of the relative error is $0.002$ at $\gamma=7$, becoming progressively smaller at larger $\gamma$). The inset shows analogous plot for the second moment $e_2(\gamma)$ that was used as an input in Eq.~(\ref{eq:e2l1}) in order to evaluate $e_4(\gamma)$.} \label{fig}
\end{figure}

The second-order differential equation (\ref{eq:e2l1}) contains exact information about the moments of the rapidity distribution. Supplemented by proper boundary (initial) conditions, Eq.~(\ref{eq:e2l1}) defines a boundary (initial) value problem that in principle can be studied studied numerically in order to obtain higher moments for intermediate values of $\gamma$ using the knowledge of $e_2(\gamma)$. However, one expects that the analytical approximations (\ref{eq:e2lass}) at $\gamma\ll 1$ and (\ref{eq:e2gass}) at $\gamma\gg 1$ taken with sufficient number of terms can well extrapolate to the regime of intermediate $\gamma$. We have confirmed this assumption for $e_2(\gamma)$ \cite{marino_exact_2019,ristivojevic_conjectures_2019,ristivojevic_excitation_2014} as well as for the case of $e_4(\gamma)$, see Fig.~\ref{fig}.

\textit{The structure of the series for $e_2$.---} To further show the usefulness of Eq.~(\ref{eq:e2l1}) [and more generally of Eq.~(\ref{eq:main1})], we can obtain the information about the series for $e_2$ of the Lieb-Liniger model in the rather complicated case $\gamma\ll 1$, since Eq.~(\ref{eq:F}) then approaches the singular limit \cite{lieb_exact_1963,tracy_ground_2016}. This can be achieved from Eq.~(\ref{eq:akoff}), which acquires a special form in the cases $k=2l+2$ and  $k=2l+4$, since its left-hand side nullifies. The right-hand side in the former case becomes a constraint on the coefficients for the series of $e_2$ and $e_{2l}$,
\begin{align}\label{eq:akoff1}
		\sum_{j=0}^{2l+2} (j-2)(j-4) a_j^{(2)} a_{2l+2-j}^{(2l)}=0.
\end{align}
For $l=1$, Eq.~(\ref{eq:akoff1}) reduces to	
\begin{align}\label{eq4}
		a_4+\frac{a_1a_3}{4a_0}=0.	
\end{align}	
Here and in the following we introduced the simplified notation by suppressing the superscript from the coefficients entering $e_2$, i.e., we use $a_j\equiv a_j^{(2)}$. In the case $k=2l+4$, Eq.~(\ref{eq:akoff}) gives another constraint, 
\begin{align}\label{eq:akoff2}
		\sum_{j=0}^{2l+4} (j-2)(j-4) a_j^{(2)} a_{2l+4-j}^{(2l)}=0.
\end{align}
Taking $l=1$, we obtain the second relation among the coefficients in $e_2$, 
\begin{align}\label{eq6}
		a_6+\frac{3a_1a_5}{8 a_0}-\frac{(a_3)^2}{16 a_0}=0.
\end{align}
The constraints (\ref{eq:akoff1}) and (\ref{eq:akoff2}) at $l>1$ in combination with Eq.~(\ref{eq:akoff}) lead to infinitely many relations among the coefficients entering the series for $e_2$. Let us illustrate how to obtain the third one. Substituting $l=2$ in Eq.~(\ref{eq:akoff2}) we obtain a sum that  involves the coefficients $a_{j}^{(4)}$ with $j=0,1,2,3,5,7,8$. Using Eq.~(\ref{eq:akoff}) we express them in terms of the sum of products of the two $a_j$ coefficients. The obtained sum of products of three $a_j$'s contains $a_2^{(2)}$ arising from the right-hand side of Eq.~(\ref{eq:akoff}). However, its overall prefactor is proportional to the left-hand side of Eq.~(\ref{eq6}) and thus nullifies. The remaining terms lead to \cite{Note1}
\begin{align}\label{eq8}
		a_8+ \frac{13  a_1 a_7}{10a_0}+\frac{7(a_1)^2 a_6}{20 a_0^2}+\frac{a_3 a_5}{2a_0} +\frac{a_1 a_3 a_4}{20 (a_0)^2}=0.
\end{align}

Equations (\ref{eq4}), (\ref{eq6}), and (\ref{eq8}) are the first three relations among the coefficients of the series for $e_2$ obtained from the general considerations based on analytic properties of the integral equation (\ref{eq:F}) and its consequence given by Eq.~(\ref{eq:e2l1}). They are in agreement with the exact numerical values for $a_j$'s \cite{marino_exact_2019,ristivojevic_conjectures_2019}. The obtained sequence of relations can be arbitrarily extended by substituting subsequently the values $l\ge 3$ in Eq.~(\ref{eq:akoff2}), followed by the repetitive use of Eqs.~(\ref{eq:akoff}) and (\ref{eq:akoff1}). The obtained relations and the subsequent ones among $a_j$'s have several special features. First, the term $a_2$ does not occur in them. Second, when multiplied by a common denominator, the summands of a particular relation have a product form $a_{j_1}a_{j_2}\cdots$ with a constant sum $j_1+j_2+\cdots$. In the relations  (\ref{eq4}), (\ref{eq6}), and (\ref{eq8}), this sum is, respectively, equal to $4$, $6$, and $8$. The second feature follows directly from Eq.~(\ref{eq:akoff}). The third feature is the possibility to express the coefficients with an even index $a_{2j}$ in terms of the coefficients with odd indices $a_1$, $a_3$,\ldots $a_{2j-1}$ and $a_0$ (which can be shown to be $a_0=1$ \cite{lieb_exact_1963}). This is obvious for Eqs.~(\ref{eq4}) and (\ref{eq6}).
The special features for the case of Eq.~(\ref{eq8}) are exemplified in Supplemental Material \cite{Note1}. Along the same lines, one can obtain further relations corresponding to $l\ge 3$.  Therefore, we have reduced the complicated problem of the series solution for $e_2$ to the problem of finding the coefficients of the series with odd indices.

\textit{Discussion.---} The moments of the rapidity distribution represent the conserved charges of Galilean-invariant integrable models in the thermodynamic limit. In this paper we have derived the relation (\ref{eq:main1}) that connects the ground-state expectation values of the consecutive conserved charges (\ref{eq:e2l}). We note that the corresponding commuting operators that have the eigenvalues (\ref{eq:e2l}) are generally unknown, apart from the first few ones in the case of the Lieb-Liniger model \cite{davies_higher_1990}.

The formalism developed in this paper expressed through Eqs.~(\ref{eq:F})-(\ref{eq:main1}) does not apply to Galilean-invariant models with attractive interactions in cases where Eq.~(\ref{eq:F}) cannot be used as a starting point. One example is the Lieb-Liniger model with attraction. It does not have well defined thermodynamic limit because the ground-state energy scales with the third power of number of particles, in contrast to the repulsive case where this scaling is linear \cite{takahashi}. We note, however, that there are models with attraction where our formalism will apply. An example is the Yang-Gaudin model of spin-$\frac{1}{2}$ fermions with attractive $\delta$-function interaction, which is described by conceptually similar equations as the Lieb-Liniger model with repulsion. The main difference arises at weak attraction where the series for the ground-state energy is with respect to the interaction parameter, which should be contrasted to Eq.~(\ref{eq:e2lass}) where the series is controlled by the square root of the interaction parameter. On the other hand, the coefficients in the series of the ground-state energy of the fermionic model at weak interaction also satisfy a number of relations, akin to Eqs.~(\ref{eq4}), (\ref{eq6}), and (\ref{eq8}) in the Lieb-Liniger case. Note that we have not specifically addressed the hyperbolic Calogero-Sutherland model, since we are not aware of works where its ground-state energy is evaluated analytically. This complicated task is beyond the scope of this paper. 

Additional interesting question is whether and how the results of this paper can be extended to account for the thermal states and moreover for more general excited states, which appear, e.g., in studies of local correlation functions \cite{pozsgay_local_2011,kormos_exact_2011}. In the case of thermal states, the Fermi step function over the rapidities in Eq.~(\ref{eq:F}) becomes a smooth Fermi function of the pseudoenergy extending the integration over the real axis, while the pseudoenergy itself satisfies a nonlinear Yang-Yang integral equation \cite{yang_thermodynamics_1969}. The problem  how to treat such equation using the method of differentiation is left for future work.

To summarize, we have shown that the moments of the rapidity distribution, equivalently the ground-state expectation values of conserved charges, in Galilean-invariant integrable models satisfy the difference-differential equation (\ref{eq:main1}). The latter implies an easy access to all higher moments once the ground-state energy, i.e., the second moment of the system is known. Knowledge of such exact results in the thermodynamic limit is generally advantageous as it can save the computation time of numerical simulations of quantum many-body systems, which only treat a limited number of particles.

The author is grateful to G.-L.~Oppo for helpful comments.


%


\onecolumngrid
\newpage
\setcounter{equation}{0}
\setcounter{figure}{0}

\renewcommand{\theequation}{S\arabic{equation}}
\renewcommand{\thepage}{S\arabic{page}}
\renewcommand{\thesection}{S\arabic{section}}
\renewcommand{\thetable}{S\arabic{table}}
\renewcommand{\thefigure}{S\arabic{figure}}

\renewcommand{\bibnumfmt}[1]{[{\normalfont S#1}]}


\setcounter{page}{1}

\begin{center}
	{\large\textbf{\TITLE}
		\\\vskip 5pt
		\normalsize{--Supplemental Material--}
		\\\vskip 8pt
	}
	
	Zoran Ristivojevic\vskip 0.5mm
	\textit{Laboratoire de Physique Th\'{e}orique, Universit\'{e} de Toulouse, CNRS, UPS, 31062 Toulouse, France}
\end{center}
\vskip 2.5pt

\textit{Appendix on the derivation of Eq.~(\ref{eq:PDE}).---} In order to derive Eq.~(\ref{eq:PDE}), it is convenient to introduce linear integral operator $\mathcal{F}[\rho(k,Q)]$ defined as the left-hand side of Eq.~(\ref{eq:F}). Performing the differentiation of $\mathcal{F}$, we directly obtain 
\begin{subequations}
\label{eq19}	
\begin{align}
	\frac{\partial}{\partial Q}\mathcal{F}[\rho] ={}& \mathcal{F}\left[ \frac{\partial\rho}{\partial Q} \right] +\frac{\rho(Q,Q)}{2\pi}[\theta'(k-Q)+\theta'(k+Q)],\\
	\frac{\partial^2}{\partial Q^2}\mathcal{F}[\rho] ={}& \mathcal{F}\left[ \frac{\partial^2\rho}{\partial Q^2} \right] +\frac{\rho'_Q(Q,Q)}{2\pi}[\theta'(k-Q)+\theta'(k+Q)]\notag\\
	&+\frac{d}{dQ}\left\{\frac{\rho(Q,Q)}{2\pi}[\theta'(k-Q)+\theta'(k+Q)]\right\}.
	\end{align}
\end{subequations}
Here we used the notation $\rho'_Q(Q,Q)=\partial\rho(k,Q)/\partial Q|_{k=Q}$. Similarly, differentiating $\mathcal{F}$ with respect to $k$, after the partial integrations we obtain
\begin{align}\label{eq20}
	\frac{\partial^2}{\partial k^2}\mathcal{F}[\rho] ={}& \mathcal{F}\left[ \frac{\partial^2\rho}{\partial k^2} \right] -\frac{\rho'_k(Q,Q)}{2\pi}[\theta'(k-Q)+\theta'(k+Q)]\notag\\
	&-\frac{\rho(Q,Q)}{2\pi}\frac{d}{dk}[\theta'(k-Q)-\theta'(k+Q)].
\end{align}
Here we have introduced $\rho'_k(Q,Q)=\partial\rho(k,Q)/\partial k|_{k=Q}$ and used the property $\rho'_k(Q,Q)=-\rho'_k(-Q,Q)$ that follows from Eq.~(\ref{eq:F}). The left-hand sides in Eqs.~(\ref{eq19}) and (\ref{eq20}) are zero, since Eq.~(\ref{eq:F}) reads $\mathcal{F}[\rho]=1/2\pi$. The linear combination of the right-hand sides then yields
\begin{align}
	\label{eq21}
	\mathcal{F}\left[\frac{\partial^2\rho}{\partial Q^2} - \frac{2}{\rho(Q,Q)}\frac{d\rho(Q,Q)}{dQ} \frac{\partial\rho}{\partial Q}-\frac{\partial^2\rho}{\partial k^2}\right]=0,
\end{align}
since $\mathcal{F}$ is a linear operator. We have also used the total derivative $d\rho(Q,Q)/dQ=\rho'_k(Q,Q)+\rho'_Q(Q,Q)$ and the parity of $\theta'(k)$. Since the integral equation (\ref{eq:F}), i.e., $\mathcal{F}[\rho]=1/2\pi$ has a unique solution \cite{lieb_exact_1963}, the Fredholm alternative theorem guarantees that Eq.~(\ref{eq21}) only has a trivial solution. This is equivalent to Eq.~(\ref{eq:PDE}), which therefore must be satisfied.

\textit{Appendix on the derivation of Eq.~(\ref{eq8}).---} Substituting $l=2$ into Eq.~(\ref{eq:akoff2}) gives
\begin{align}\label{eq8'}
	8 a_0 a_8^{(4)}+3a_1 a_7^{(4)}-a_3 a_5^{(4)}+3a_5 a_3^{(4)}+8 a_6 a_2^{(4)}\notag\\
	+15 a_7 a_1^{(4)}+24 a_8a_0^{(4)}=0.
\end{align}
Equation (\ref{eq:akoff}) at $l=1$ leads to $a_k^{(4)}=\frac{6}{(4-k)(6-k)}\sum_{j=0}^{k} (j-2)(j-4)a_j a_{k-j}$. Substituting the latter into Eq.~(\ref{eq8'}) leads to a linear combination of terms of the form $a_{j_1}a_{j_2}a_{j_3}$, where $j_1+j_2+j_3=8$. The coefficient $a_2$ in the combination arises with the coefficient $6[16a_0a_6+6a_1a_5-(a_3)^2]$, which is zero due to  Eq.~(\ref{eq6}). The remainder then gives Eq.~(\ref{eq8}). Expressing $a_4$ and $a_6$ in Eq.~(\ref{eq8}) obtained from Eqs.~(\ref{eq4}) and (\ref{eq6}), we eventually obtain
\begin{align}\label{eqlast}
	a_8=-\frac{13 a_1a_7}{10a_0}-\frac{a_3a_5}{2a_0}+\frac{21 (a_1)^3 a_5}{160(a_0)^3}-\frac{3(a_1)^2(a_3)^2}{320(a_0)^3}.
\end{align}
Equation (\ref{eqlast}) is an expression for $a_8$ in terms of the coefficients with odd indices and $a_0$ (which is actually $a_0=1)$.

\end{document}